\documentclass[12pt]{article}%
\usepackage{amsmath,latexsym}
\usepackage{graphicx}
\usepackage{amsmath}
\usepackage{amsfonts}
\usepackage{amssymb}%
\begin{document}

\title{{A note on wormholes as compact stellar objects}}
   \author{
Peter K. F. Kuhfittig*\\  \footnote{kuhfitti@msoe.edu}
 \small Department of Mathematics, Milwaukee School of
Engineering,\\
\small Milwaukee, Wisconsin 53202-3109, USA}

\date{}
 \maketitle

\begin{abstract}\noindent
A Morris-Thorne wormhole can only be sustained
by violating the null energy condition,
thereby requiring the existence of ``exotic
matter,"a requirement that many researchers
consider unphysical.  Using a mostly
qualitative approach, it is proposed in
this note that the basic problems can be
overcome by purely relativistic considerations.
The implication is that a Morris-Thorne
wormhole is likely to be a compact stellar
object.

\end{abstract}

\section{Introduction}\label{E:introduction}

It is well known that a Morris-Thorne
wormhole can only be held open by violating
the null energy condition (NEC), calling
for the existence of exotic matter at or
near the throat.  While the need for exotic
mater is rather problematical, it is not a
conceptual problem, as we know from the
Casimir effect \cite{MT88}.  In other
words, exotic matter can be made in the
laboratory.  An open question is whether
enough could be produced locally to sustain
a macroscopic traversable wormhole.  Many
researchers consider any wormhole solution
unphysical if exotic matter cannot be
avoided.  The purpose of this note is to
reexamine some of these requirements.  We
are primarily interested in qualitative
results.

\section{Wormhole structure}

We start this section with the following
line element, using units in which $c=G=1$
\cite{MTW}:
\begin{equation*}
  ds^{2}=-e^{2\Phi(r)}dt^{2}+\frac{dr^2}
  {1-\frac{2m(r)}{r}}
  +r^{2}(d\theta^{2}+\text{sin}^{2}\theta\,
  d\phi^{2}),\quad r\le R
  \end{equation*}
  \begin{equation}\label{E:line1}
  =-\left(1-\frac{2M}{r}\right)dt^2
  +\frac{dr^2}{1-\frac{2M}{r}}
  +r^{2}(d\theta^{2}+\text{sin}^{2}\theta\,
  d\phi^{2}), \quad r>R.
\end{equation}
Here $m(r)$ is the effective mass inside
radius $r$ and $M$ is the mass of a star of
radius $R$ as seen by a distant observer.
If $\rho(r)$ is the energy density, then
the total mass-energy inside radius $r$
is given by
\begin{equation}
   m(r)=\int^r_04\pi r^2\rho(r)\,dr,
   \quad m(0)=0.
\end{equation}

The line element of a Morris-Thorne
wormhole is \cite{MT88}:
\begin{equation}\label{E:line}
  ds^{2}=-e^{2\Phi(r)}dt^{2}
  +\frac{dr^2}{1-\frac{b(r)}{r}}
  +r^{2}(d\theta^{2}+\text{sin}^{2}\theta\,
  d\phi^{2}),
\end{equation}
using units in which $c=G=1$.  Here
$\Phi=\Phi(r)$ is called the \emph{redshift
function}, which must be everywhere finite
to prevent an event horizon.  The function
$b=b(r)$ is called the \emph{shape function}
since it determines the spatial shape of the
wormhole when viewed, for example, in an
embedding diagram \cite{MT88}.  The spherical
surface $r=r_0$ is called the \emph{throat}
of the wormhole, where $b(r_0)=r_0$.  The
shape function must also meet the requirement
$b'(r_0)<1$, called the \emph{flare-out
condition}, while $b(r)<r$ for $r>r_0$.  We
also require that $b'(r_0)>0$.   A final
requirement is asymptotic flatness:
$\text{lim}_{r\rightarrow
\infty}\Phi(r)=0$ and $\text{lim}_{r\rightarrow
\infty}b(r)/r=0$.

The flare-out condition can only be met by
violating the null energy condition (NEC)
which states that
\begin{equation}
  T_{\alpha\beta}k^{\alpha}k^{\beta}\ge 0
\end{equation}
for all null vectors $k^{\alpha}$, where
$T_{\alpha\beta}$ is the energy-momentum
tensor.  Matter that violates the NEC is
called ``exotic" in Ref. \cite{MT88}.  In
particular, for the outgoing null vector
$(1,1,0,0)$, the violation has the form
\begin{equation}\label{E:violation}
   T_{\alpha\beta}k^{\alpha}k^{\beta}=
   \rho +p_r<0.
\end{equation}
Here $T^t_{\phantom{tt}t}=-\rho$ is the
energy density, $T^r_{\phantom{rr}r}= p_r$
is the radial pressure, and
$T^\theta_{\phantom{\theta\theta}\theta}=
T^\phi_{\phantom{\phi\phi}\phi}=p_t$ is
the lateral (transverse) pressure.  Before
continuing, let us list the Einstein field
equations:

\begin{equation}\label{E:Einstein1}
  \rho(r)=\frac{b'}{8\pi r^2},
\end{equation}
\begin{equation}\label{E:Einstein2}
   p_r(r)=\frac{1}{8\pi}\left[-\frac{b}{r^3}+
   2\left(1-\frac{b}{r}\right)\frac{\Phi'}{r}
   \right],
\end{equation}
and
\begin{equation}\label{E:Einstein3}
   p_t(r)=\frac{1}{8\pi}\left(1-\frac{b}{r}\right)
   \left[\Phi''-\frac{b'r-b}{2r(r-b)}\Phi'
   +(\Phi')^2+\frac{\Phi'}{r}-
   \frac{b'r-b}{2r^2(r-b)}\right].
\end{equation}

To see the connection to the flare-out
condition at the throat, observe that from
Eqs. (\ref{E:violation}), (\ref{E:Einstein1}),
and (\ref{E:Einstein2}), we have
\begin{equation}\label{E:exotic}
   8\pi[\rho(r_0)+p_r(r_0)]=\frac{r_0b'(r_0)
   -b(r_0)}{r_0^3}<0   
\end{equation}
since $b(r_0)=r_0$.  Given that the radial 
tension $\tau(r)$ is the negative of 
$p_r(r)$, Eq. (\ref{E:violation}) can be 
written
\begin{equation}
   \tau-\rho c^2>0,
\end{equation}
temporarily reintroducing $c$.  The last
inequality has given rise to the designation
``exotic matter" since $\tau>\rho c^2$
implies that there is an enormous radial
tension at the throat.  This problem has
been discussed extensively in Refs.
\cite{pK13, pK20, pK21, pK22a, pK22b}.

\section{The flare-out condition}

It follows from Eq. (\ref{E:Einstein1}) that
\begin{equation}
   b(r)=r_0+\int^r_{r_0}8\pi r^2\rho(r)\,dr,
\end{equation}
confirming that $b(r_0)=r_0$, now viewed as
an initial condition.  The physical
significance of this condition will be
examined more closely in this note.  Observe
also that
\begin{equation}\label{E:mass}
   b(r)=2 m(r).
\end{equation}

As noted in the Introduction, we are primarily
interested in qualitative results.  To that
end, we first recall that $\rho(r)$ is likely
to be very small in geometrized units.  So it
follows that
\begin{equation}
    b'(r_0)=8\pi r_0^2\rho(r_0)<1;
\end{equation}
so the flare-out condition has been met.  To
show that the assumption regarding $\rho$ is
realistic, suppose we try $\rho(r_0)=
10^{-2}\text{m}^{-2}$.  Then
\[
   \rho(r_0)=10^{-2}\frac{c^2}{G}=1.35
   \times 10^{25}\frac{\text{kg}}{\,\text{m}^3}.
\]
Given that nuclear matter has a density of
$10^{18}$\,kg/$\text{m}^3$, our choice of
$\rho$ could be even smaller than $10^{-2}
\,\text{m}^{-2}$.

\section{The condition $b(r_0)=r_0$}
We know from Eq. (\ref{E:mass}) that
$\frac{1}{2}b(r)$ is the effective mass
inside radius $r$.  Since $r=r_0$ is the
throat of the wormhole, it follows from
the definition of throat that the interior
$r<r_0$ is outside the wormhole spacetime.
So $\frac{1}{2}b(r_0)=m(r_0)$ must be the
mass of the interior $r<r_0$.  This can
be compared to a thin-shell wormhole from
a Schwarzschild black hole \cite{mV89}.
The radius of the throat necessarily
exceeds the radius of the event horizon.
So while the black hole is not part of
the wormhole spacetime, it helps produce
the necessary gravitational field.
Similarly, according to Ref. \cite{pK13},
the existence of a massive core of quark
matter at the center of a neutron star
could give rise to a wormhole.  (See
also Ref. \cite{DFKK}.)

In general, then, the mass of the interior
is $\frac{1}{2}b(r_0)=\frac{1}{2}r_0$.  At
first glance this appears to be impossible:
for example, in geometrized units, the mass
of the Earth is 0.44 cm and the mass of the
sun is 1.5 km; both are very much less than
the corresponding radii.  Although seemingly
absurd, we can accept the condition $b(r_0)
=r_0$ by appealing to exotic matter, not
because this solves the problem but because
we can now claim that a sufficiently far
advanced civilization may be able to handle
the resulting technical difficulties.  An
obvious alternative is to take into account
certain relativistic effects in order to
satisfy the condition $m(r_0)=\frac{1}{2}r_0$.

Since we are primarily interested is
qualitative results, let us assume that
$\rho$ is constant, as in the original
Schwarzschild interior solution.  Since 
$m(r)$ has units of length, it follows 
from line element (\ref{E:line1}) that 
the element of volume is given by the
relativistic form
\begin{equation}
   dV(r)=4\pi r^2\frac{1}
   {\sqrt{1-\frac{2 m(r)}{r}}}\,dr.
\end{equation}
Recalling that $m(r)$ is the effective 
mass inside radius $r$, we get 
\begin{equation}
   \frac{2m(r)}{r}=2\cdot\frac{4}{3}
   \frac{\pi r^3}{r}\rho=\frac{8}{3}
   \pi r^2\rho,
\end{equation}
since $\rho$ is a constant.  It follows 
that 
\begin{equation}
   dV(r)=4\pi r^2\frac{1}
   {\sqrt{1-\frac{8}{3}\pi r^2\rho}}\,dr
\end{equation}
and 
\begin{equation}
  V(r)=\int^r_04\pi r^2\frac{1}
   {\sqrt{1-\frac{8}{3}\pi r^2\rho}}\,dr\\
   =4\pi\left[\frac{\text{sin}^{-1}
   \left(r\sqrt{\frac{8}{3}\pi\rho}\right)}
   {2\left(\frac{8}{3}\pi\rho\right)^{3/2}}
   -\frac{r\sqrt{1-\frac{8}{3}\pi r^2\rho}}
   {2\left(\frac{8}{3}\pi\rho\right)}\right].
\end{equation}

Next, we observe that
\[
  \text{lim}_{x\rightarrow 0}
  \frac{\text{sin}^{-1}x}{x}=
  \text{lim}_{x\rightarrow 0}
  \frac{\frac{1}{\sqrt{1-x^2}}}{1}=1
\]
by L'Hospital's rule.  So for small values
of $x$, $\text{sin}^{-1}x\approx x$ and
for $\rho$ sufficiently small, it then
follows that
\begin{multline}
  V(r)\approx
   4\pi\left[\frac{
   r\sqrt{\frac{8}{3}\pi\rho}}
   {2\left(\frac{8}{3}\pi\rho\right)^{3/2}}
   -\frac{r\sqrt{1-\frac{8}{3}\pi r^2\rho}}
   {2\left(\frac{8}{3}\pi\rho\right)}\right]
   =4\pi\left[
   \frac{r}{\frac{16}{3}\pi\rho}
   -\frac{r\sqrt{1-\frac{8}{3}\pi r^2\rho}}
   {\frac{16}{3}\pi\rho}\right]\\
   =4\pi\frac{r\left(1-\sqrt{1-
   \frac{8}{3}\pi r^2\rho}\right)}
   {\frac{16}{3}\pi\rho}
\end{multline}
and
\begin{equation}
   V(r_0)=4\pi\frac{r_0\left(1-\sqrt{1-
   \frac{8}{3}\pi r_0^2\rho}\right)}
   {\frac{16}{3}\pi\rho},
\end{equation}
the total volume inside $r=r_0$.  Since 
we are primarily interested in qualitative 
results, we can choose a small $\rho$ so 
that $\frac{8}{3}\pi r_0^2\rho$ is close 
to unity.  So the effective mass (measured 
in meters) is given by  
\begin{equation}
   M(r_0)=\frac{V(r_0)}{r_0^2}=
   4\pi\frac{r_0}{\frac{16}{3}\pi\rho r_0^2}
    =\frac{1}{\frac{4}{3}\rho r_0}
    \gg\frac{1}{2}b(r_0)
\end{equation}
 since $\rho\ll 1$.  The relativistic mass 
 $M(r_0)$ could therefore be large enough 
 to meet the condition $M(r_0)=\frac{1}{2}
 r_0$.
 
 \emph{Remark:} Since we are requiring that 
 $\rho \approx 1/(\frac{8}{3}\pi r_0^2)$, 
 our conclusion is valid only for relatively 
 large throat sizes.

\section{Summary}

Our mostly qualitative approach has
shown that the boundary condition for
a wormhole, $b(r_0)=r_0$, which has
proved to be highly problematical,
can be met by means of purely
relativistic considerations.  So the
wormhole structure would have to be
extremely massive to begin with.  With
the neutron-star example in mind, it
seems quite likely that a traversable
Morris-Thorne wormhole would be a
compact stellar object.  This conclusion
is consistent with the proposed search
for wormholes by means of gravitational
lensing.  This technique lends itself \
to a direct detection of photon spheres,
but these would only exist for compact
stellar objects.

\end{document}